\documentclass[runningheads,fleqn]{svmult}
\usepackage{color} 
\usepackage{makeidx}   % allows index generation
\usepackage{graphicx}  % standard LaTeX graphics tool
\usepackage{subeqnar}  % subnumbers individual equations
\usepackage{multicol}  % used for the two-column index
\usepackage{taphys}    % flushleft layout of captions etc.
\makeindex             % used for the subject index

\begin{document}
\title*{The Role of Contacts in Molecular Electronics{\footnote{%
This chapter is based on two invited talks (from GC and FG) given at the
DPG Spring Meeting (March 2002; Regensburg, Germany)}}}
\toctitle{The Role of Contacts in Molecular Electronics
\protect\newline }
\titlerunning{The role of contacts in molecular electronics}
\author{Gianaurelio Cuniberti\inst{1}\thanks{%
Corresponding author: \texttt{cunibert@mpipks-dresden.mpg.de}}
\and Frank Gro{\ss}mann\inst{2}%
\and Rafael Guti{\'e}rrez\inst{2}%
}
\authorrunning{G. Cuniberti \textit{et al.}}
\institute{Max Planck Institute for the Physics of Complex Systems, \\
D-01187 Dresden, Germany
\and Institute for Theoretical Physics, Technical University of Dresden, \\
D-01062 Dresden, Germany}

\maketitle              % typesets the title of the contribution

\setlength{\parindent}{0.pt}
\newcommand{\dadefinire}[1]{\textcolor{red}{{{\textbf{#1}}}}} 
\newcommand{\blue}[1]{{\textcolor{blue}{\scriptsize \textsc{\textbf{#1}}}}}
\newcommand{\newpageop}{\newpage}
\newcommand{\mcal}[1]{\mathcal{#1}}
\newcommand{\mcalE}{\eta}
\makeatletter
% define det function as \det
\def\det{\mathop{\operator@font det}\nolimits}
% define Re function as \Re
\def\Re{\mathop{\operator@font Re}\nolimits}
% define Im function as \Im
\def\Im{\mathop{\operator@font Im}\nolimits}
\makeatother
\def\ie{\textit{i.e.}}
\def\eg{e.g.}
\def\etal{\textit{et al.}}
\def\ee{{\rm e}}
\def\ii{{\rm i}}
\renewcommand{\vec}[1] {{\mathbf{#1}}}
\newcommand{\vt} {\vartheta}
\newcommand{\bea} {\begin{eqnarray}}
\newcommand{\eea} {\end{eqnarray}}
\newcommand{\beann} {\begin{eqnarray*}}
\newcommand{\eeann} {\end{eqnarray*}}
\newcommand{\labs} {\left\vert}
\newcommand{\rabs} {\right\vert}
\newcommand{\lsb} {\left[}
\newcommand{\rsb} {\right]}
\newcommand{\lrb} {\left(}
\newcommand{\rrb} {\right)}
\newcommand{\lcb} {\left\{}
\newcommand{\rcb} {\right\}}
\newcommand{\lab} {\left\langle}
\newcommand{\rab} {\right\rangle}
\newcommand{\ve} {\varepsilon}
\newcommand{\ds} {\displaystyle}

\begin{picture}(1,1)
\put(-110,180){
\begin{minipage}{19cm}
{\small
In \textit{Advances in Solid State Physics}, Vol. 42, B. Kramer Ed. Springer (2002); ISBN:
3-540-42907-7.
\\
Preprint ref: \texttt{mpi-pks/0201005}
\mbox{}
\hfill {March 12, 2002}
}
\end{minipage}
}
\end{picture}

\begin{abstract}
Molecular electronic devices are the upmost destiny of the miniaturization
trend of electronic components. Although not yet reproducible on large scale,
molecular devices are since recently subject of intense studies both
experimentally and theoretically, which agree in pointing out the extreme
sensitivity of such devices on the nature and quality of the contacts.
This chapter intends to provide a general theoretical framework for modelling electronic transport at the molecular scale by describing the implementation of a hybrid method based on Green function theory and density functional algorithms.
In order to show the presence of contact-dependent features in the molecular conductance, 
we discuss three archetypal molecular devices, which are intended to focus
on the importance of the different sub-parts of a molecular two-terminal setup.
\end{abstract}

\section{Introduction}
\index{molecular electronics}
\index{molecular devices}
\index{lead-molecule contact}
The incessant development of single molecule techniques is forcing a paradigm
shift in the many neighboring branches of nano-sciences. This process does not
exclude the modelling and design of electronic devices.
Novel fabrication methods that create metallic contacts to a small number of conjugated organic molecules allow the study of the basic transport mechanism of these systems and will provide direction for the potential development of molecular-scale electronic systems~\cite{Reed99}.
The concept is now realized for individual components, but the economic fabrication of complete circuits at the molecular level remains challenging because of the difficulty of connecting molecules to one another. A possible solution to this problem is `mono-molecular' electronics, in which a single molecule will integrate the elementary functions and interconnections required for computation~\cite{JGA00}.
Indeed, the primary problems facing the molecular electronics designer are measuring and predicting electron transport. That is due to the fact that
molecular electronics is strongly dependent on the quality and nature
of the contacts~\cite{CFR01a}. Ideally, these contacts should be ohmic so that any non-linearity in the conductivity of the wire can be correctly attributed and studied. They must also be low in resistance to ensure that the properties measured are those of the molecule and not those of the molecule-contact interface. Moreover, the medium surrounding and supporting the molecule must be several orders of magnitude more insulating than the molecule itself because the contact area of the support with the electrical contacts is often much greater than that between the electrical contacts and the molecule~\cite{Hipps01}.

Nevertheless, the contact problem can be turned into a challenge.
Even with the intrinsic barrier that the contacts represent, barriers can be strategically used to favor the design of specific devices~\cite{SDLCD01}. However, this requires a more detail account of the atomic structure of the interface.
Green function and density functional theories~\cite{CDeMV00+DS01} are the typical instruments to characterize transport through single molecules clamped between two metallic contacts. These very same instruments may even be adopted for calculating electromechanical switch behaviors~\cite{GFCGRS02+CGFGRS02} and current-induced forces~\cite{DVPL02} in molecular structures.

In this chapter, after a brief overview on charge transport on the molecular
scale (Sect.~\ref{sect:ctms}), we provide, in Sect.~\ref{sect:method}, a general theoretical hybrid method based on Green function theory and density functional
theory (DFT)-based algorithms.
In order to show the presence of contact-dependent features in the molecular
conductance, we introduce, in Sect.~\ref{sect:application}, three model molecular devices.
The first is a sodium wire (Par.~\ref{subsect:sodium}), where the role of contacts for a molecular bridge emerges clearly.
However, the quality of contacts is not the only source of alteration of the molecular conductance. In the Par.~\ref{subsect:cnt}, we show the peculiar effect that carbon nanotube leads might have on a contacted molecule. Finally,
in the Par.~\ref{subsect:purecarbon},
a pure carbon device, consisting of two carbon nanotube leads grasping a C$_{60}$ molecule is studied in a parameter free DFT calculation.

\section{Charge Transport on the Molecular Scale}\label{sect:ctms}

\begin{figure}[t]
\centerline{\includegraphics[width=.60\textwidth]{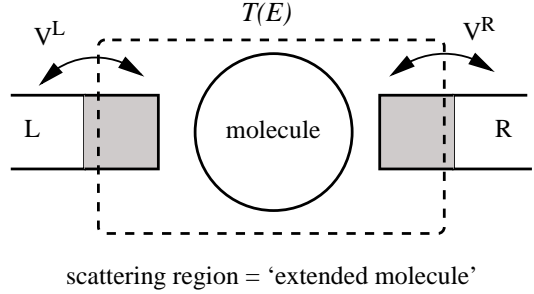}}
\caption[]{\label{fig:schem}Schematic representation of a two-terminal device.
The scattering region (enclosed in
the dashed-line frame) with transmission
probability $T(E)$ is connected to semi-infinite left (L) and right (R) leads
which end into
electronic reservoirs (not shown) at chemical potentials
$\mu_{\mathrm{L}},\mu_ {\mathrm{R}}$. By applying a
small potential difference electronic transport will occur. The scattering
region or molecule
may include in general parts of the leads (shaded areas). This is necessary
for the description
of real systems,
where the surface topology may be modified as a result of
relaxation or reconstruction processes. This may introduce additional
scattering due, \eg, to
surface states.}
\end{figure}

In mesoscopic electron transport, many interesting interference related
and quantization effects have been found in the past 20 years~\cite{Kramer94}.
Much of the fundamental theory for mesoscopic systems
can be taken over to the description
of molecular scale conductance calculations. In both realms,
a formulation that includes interference effects due to phase coherence
as well as geometrical
effects is needed. It was originally developed by
\index{Landauer approach}
Landauer~\cite{Landauer57} for a
two-terminal geometry as
displayed in Fig.~\ref{fig:schem}.\
and further extended by
B{\"u}ttiker~\cite{Buettiker88} to the multi-terminal case.
The essential idea of the Landauer formulation is to relate the
conductance to an elastic scattering problem and, ultimately, to transmission
probabilities. The
simplest way to derive this relationship is to consider a `molecular' region
connected to two
ballistic leads, which are connected to electronic reservoirs at the chemical
potentials
$\mu_{\mathrm{L}},\mu_{\mathrm{R}}$, see Fig.~\ref{fig:schem}.\ It is assumed
that electrons entering the reservoirs do
completely lose their phase coherence. As stated in Ref.~\cite{Nikolic01}
assuming semi-infinite
leads is enough to warrant that no electron exiting the scattering region will
reenter it
with the same phase, so that an explicit modeling of the reservoirs is not
necessary.
In equilibrium $\mu_{\mathrm{L}}=\mu_{\mathrm{R}}$, but if an
infinitesimal voltage $eV=\mu_{\mathrm{L}}-\mu_{\mathrm{R}}$
is applied a
non-equilibrium situation is induced and a current will flow.
The scattering region is
characterized by the energy-dependent transmission coefficient $T(E)$. In the
zero-temperature,
linear response ($eV\to 0$)
regime it is found~\cite{Datta99} that the proportionality relation,
\begin{eqnarray}
\label{eq:1}
g=\frac{e^2}{\pi\hbar} T(E_{\mathrm{F}}), \; \; \; \; \; T=\sum_{m,n=1}^{N_
{\mathrm{F}}} |t_{mn}|^2
\end{eqnarray}
holds. $E_{\mathrm{F}}$ is the Fermi energy of the whole system in equilibrium
and the transmission
amplitudes $t_{mn}$ describe the scattering of one electron from channel $n$
in the left lead
to channel $m$ in the right lead. They can be extracted, \eg, from the
scattering matrix.
The sums run over all $N_{\mathrm{F}}$ open channels at the Fermi level (whose
number
is assumed to be equal on both sides). Channels (transverse modes) appear
due to the finite cross section of the leads which induces quantization
of the electronic
states perpendicular to the direction of current transport.
For the special case of ideal transmission,
i.\ e.\ $\sum_n|t_{mn}|^2=1 \; \forall m$, the conductance is
simply proportional to $N_{\mathrm{F}}$ with the von
Klitzing conductance quantum \index{conductance quantum}
$g_{\mathrm{K}}=e^2/(\pi\hbar)$ as the proportionality factor.
Thus, $g/g_{\mathrm{K}}$
shows unit steps as a function of $N_{\mathrm{F}}$.
This is the well-known fact of conductance quantization~\cite{Datta99},
shown experimentally by van Wees \etal~\cite{vWvHBWKvdMF88} and Wharam \etal~\cite{WTNPAFHPRJ88}.

In molecular conductance, in the case of strong coupling, it is the
electronic structure of the molecule influenced by the leads that
determines the transmission properties which in turn play the decisive role
for electron transport. The calculation
of the coupling to the leads together with the calculation of the electronic
structure will be dealt with in the following using DFT-based methodology.

\section{Method}\label{sect:method}\index{Green function techniques}
For a general scattering region where inelastic effects are included,
Meir and Wingreen~\cite{MW92} used non-equilibrium Green functions
to derive an expression for the current which reduces to Eq.~(\ref{eq:1})
in the elastic case.
An advantage of their derivation is that an explicit
connection to the Green function $\vec{G}$ of the \textit{scattering region}
dressed by the presence of the leads is established. The latter are
introduced as self-energy
\index{lead self-energy}
corrections into the bare `molecular' Green function
${\vec{G}^{\phantom{1}}\! \!}^{-1} = {\vec{G}^{\rm M}}^{-1} +
\vec{\Sigma^{\rm L} + \Sigma^{\rm R}}$,
The result for the
transmission probability is then given by~\cite{MW92}
\begin{eqnarray}
\label{eq:2}
T (E) = 4 \; {\rm Tr} \lcb \vec{\Delta}^{\rm L} (E) {\vec{G}} (E)
\vec{\Delta}^{\rm R} (E) \vec{G}^\dagger (E) \rcb ,
\end{eqnarray}
where $\vec{\Delta}^{\mathrm{L}},\vec{\Delta}^{\mathrm{R}}$ are the spectral density describing the coupling of the scattering region
to the $\alpha$(=L,R)-lead given by
\beann
\vec{\Delta}_\alpha
= \frac \ii 2 \lrb \vec{\Sigma}_\alpha^{\phantom{\dagger}}\lrb E+\ii 0^+ \rrb - \vec{\Sigma}_\alpha^\dagger \lrb E + \ii 0^+ \rrb \rrb,
\eeann
and the trace is to be taken over states in the scattering
region.
The Green function $ {\vec{G}^{\mathrm{M}}}$ is in general defined as the inverse operator
$(E+\ii 0^+ -\vec{H}^{\mathrm{M}})^{-1}$, for some suitable `molecular' Hamiltonian $\vec{H}^{\mathrm{M}}$. Similar expressions have been derived by Fisher and Lee
and Todorov, Briggs and Sutton~\cite{FL81+TBS93}. In a seminal paper Szafer and Stone have
derived the Landauer result, Eq.~(\ref{eq:1}) from Kubo's linear response theory~\cite{SS88}.

As mentioned above, only components of the Green function in the Hilbert subspace associated with
the scattering region, which we will denote as `the molecule' from now on to keep in mind that
transport through molecular scale systems is the main issue to be addressed here, are needed.
Notice, however, that the molecule may also include some atoms belonging to the leads, see Fig.~\ref{fig:schem}.
This will be the case when investigating \textit{real} systems, where the surface atomic structure
of the leads is explicitly taken into account (clean surfaces are usually energetically unstable,
so that upon structural relaxation the surface topology may be modified and this will introduce
additional scattering). From the full Green function of the open (infinite) system consisting of the leads plus the molecule it is possible to extract
$ {\vec{G}}$ using projector operator techniques~\cite{PSRB96}. In order to do
so, one partitions the whole system into
three components as shown in Fig.~\ref{fig:schem}., where a left electrode, the molecule,
and a right electrode are depicted. The full associated Hamiltonian matrix
(in a suitable basis representation) can then be formally written as
\begin{eqnarray}
\vec{H} = \left (
\begin{array}{c c c}
\vec{H}^{\mathrm{L}} & \vec{V}^{\mathrm{L,M}} & \mathbf{0}\\
{\vec{V}^{\mathrm{L,M}}}^{\dagger} & \vec{H}^{\mathrm{M}} & \vec{V}^{\mathrm{R,M}} \\
\mathbf{0}& {\vec{V}^{\mathrm{R,M}}}^{\dagger} & \vec{H}^{\mathrm{R}}
\end{array}
\right ).
\end{eqnarray}
The matrices $\vec{V}^{\mathrm{L,M}},\vec{V}^{\mathrm{R,M}}$ couple atoms belonging to the left(right) leads to the molecule,
and it has been assumed that no direct lead-lead coupling exists.
Notice that $\vec{H}^{\mathrm{L(R)}}$ are infinite dimensional sub-matrices.
By means of a operator projecting onto the `molecular' subspace, we can write
the resulting $M$-dimensional matrix equation as:
\begin{eqnarray}
\lrb \phantom{|^|}\!\!\! z { \vec{S}^{\mathrm{M}}} -{ \vec{H}^{\mathrm{M}}} -{ \vec{\Sigma}^{\mathrm{L}}} \lrb z \rrb -{ \vec{\Sigma}^{\mathrm{R}}} \lrb z \rrb \rrb \vec{G} \lrb z \rrb = { \mathbf{1}}, \; \;
\; z=E + \ii 0^+,
\end{eqnarray}
where $\vec{S}^{\mathrm{M}}$ is the overlap matrix for the general case of a non-orthogonal basis set.
The energy-dependent
self-energies $\vec{\Sigma}^{\mathrm{L}},\vec{\Sigma}^{\mathrm{R}}$ include the coupling to the leads as well as information
on the electronic structure of the leads. For the $\alpha$-lead, they are given by
\begin{eqnarray}
\vec{\Sigma}^\alpha \lrb z \rrb = \lrb z\, {\vec{S}^{\alpha,\mathrm{M}}} - {\vec{V}^{\alpha,\mathrm{M}}} \rrb^{\dagger}
{\mathcal{G}}^\alpha \lrb z \rrb \lrb z\, \vec{S}^{\alpha,\mathrm{M}}- {\vec{V}^{\alpha,\mathrm{M}}}
\rrb .
\end{eqnarray}
The matrix $\vec{S}^\alpha$ is the overlap matrix
element between molecule atoms and the $\alpha$-lead atoms and $\mathcal{G}^\alpha (z)$
is the $\alpha$-lead surface Green functions. Since the coupling matrices are in general
short-ranged
they will
eliminate all contributions coming from atoms other
than those closest to the molecule. Hence, only
surface Green functions are usually needed.

We would like to
stress that Eqs.~(\ref{eq:1}) and (\ref{eq:2}) are only valid in the case that
inelastic processes
in the scattering region can be
completely neglected. Otherwise no simple relationship between conductance and
transmission can be obtained.
A typical example where electron-electron interactions are decisive
are quantum dots. There, the scattering region is weakly
coupled to the leads so that
the coupling-induced level broadening will be much smaller than the
charging energy. Hence,
electron interaction effects leading, \eg, to Coulomb blockade phenomena
should be included in the
description of quantum transport~\cite{GD92-asi}.

At this point we are led to the issue of characterizing the electronic structure of the
molecule as well as of the leads. If we only focus on the essential physics, some kind of model
Hamiltonians can be used~\cite{MKR94b}. However, if {\it real} situations are addressed where the knowledge of the
detailed electronic structure is important, the use of more realistic computational
schemes is unavoidable. From the point of view of electronic structure calculations three classes
of approaches have been implemented for quantum transport calculations:
\begin{itemize}
\item[(i)] Semiempirical or empirical tight-binding (TB) schemes, \eg~(extended) H{\"u}ckel Hamiltonians, where the matrix elements are fitted to experiments or to first-principle calculations~\cite{GBH97+DTHRHK97+MJ97+EK98a+PS99+CBLC96}.
\index{tight-binding Hamiltonians}
\item[(ii)] First-principles or \textit{ab initio} approaches like Hartree-Fock
and (DFT)~\cite{LA00a+YRGMR99+PPJLV01+SLC01+TGW01+DGD01}.
\index{\textit{ab initio} methods}
\index{density functional theory (DFT)}
\item[(iii)] Schemes which combine some elements of points (i) and (ii) in first-principles
parametrized tight-binding Hamiltonians as it is the case for TB-DFT~\cite{GFCGRS02+CGFGRS02,GGS01a,GGKS01,TFPGGS02}
methods.
\end{itemize}

Concerning the last class mentioned above,
a computational scheme has been developed in Ref.~\cite{GGKS01}
which combines a DFT-parametrized TB
approach
with the Landauer formalism to study the
electronic transport properties of sodium atomic
chains~\cite{GGS01a}, small sodium clusters~\cite{GGKS01}, carbon-based molecular
junctions~\cite{GFCGRS02+CGFGRS02} as well as to simulate Scanning-Tunneling-Spectroscopy experiments
on organic molecules~\cite{TFPGGS02}. The TB-DFT scheme relies on a representation of the
electronic eigenstates of the system within
a non-orthogonal localized basis set, usually taken as a valence basis.
The many-body Hamiltonian is then approximately represented by
a two-center tight-binding Hamiltonian. The matrix
elements, however, are calculated numerically, avoiding the introduction of
empirical parameters as in conventional TB approaches.
We will now discuss some of the applications of this
combined scheme.

\section{Applications to Molecular Devices}\label{sect:application}
\subsection{Focusing on the Bridge Molecule: Sodium Wires}\label{subsect:sodium}
\index{sodium wires}
\index{molecular wires}
\begin{figure}[t]
\centerline{\includegraphics[width=.99\textwidth,height=.40\textwidth]{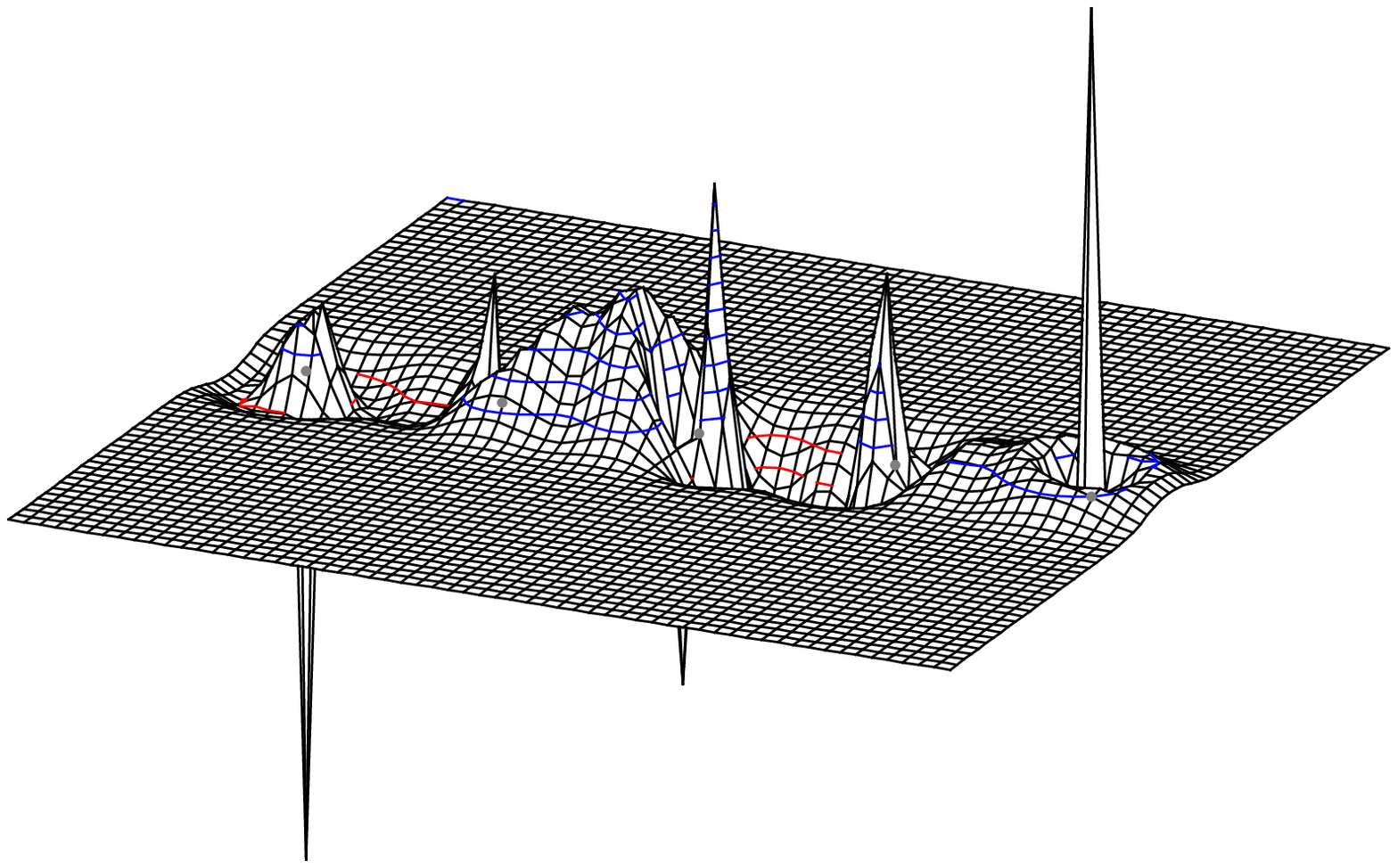}%
}
\caption[]{\label{fig:sodiumsketch}A five sodium atom isolated wire. The HOMO density is plotted in arbitrary units.}
\end{figure}
In this section, we review the numerical results for the resistance ${\cal R}=1/{g}$
of sodium atomic wires as a function of the electrode-wire separation $d$
and of the wire length~\cite{GGS01a}. The bond length in the wires was fixed at 6.00 $a_{\mathrm{B}}$, which
approximately corresponds to the equilibrium distance of a Na-dimer
($d_{\rm eq}$=5.67 $a_{\mathrm{B}}$). For wires with more than four atoms dimerization of the wire
is expected due to a Peierls transition. Such effects will not
be considered here.

\begin{figure}[t]
\centerline{\includegraphics[width=.90\textwidth]{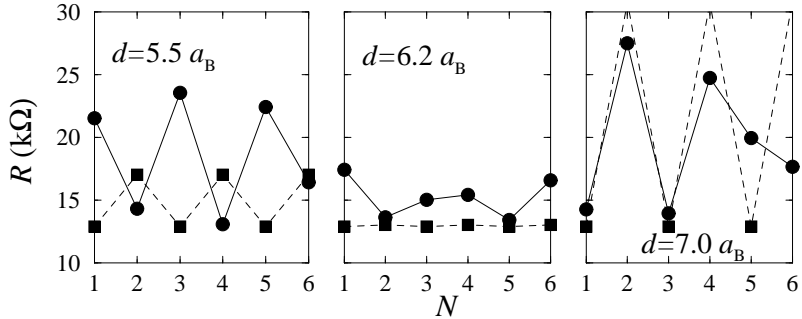}}
\caption[]{\label{fig:sod1}Dependence of the resistance on the length
of the atomic sodium wire for different electrode-wire separations. Dashed
lines (connecting the squares) correspond to a resistance calculated with only
the 3s-valence orbitals in the wire.}
\end{figure}

\begin{figure}[b]
\centerline{\includegraphics[width=.90\textwidth]{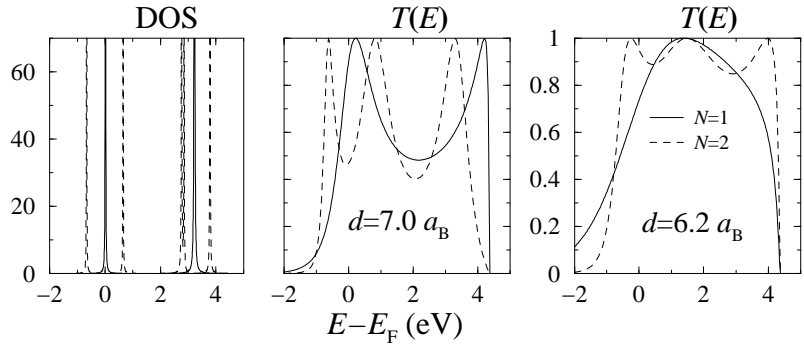}}
\caption[]{\label{fig:sod2}The transmission coefficient as a function of energy for one and two atoms between the electrodes for two different electrode-wire separations. The left panel shows the DOS of the isolated wires. Only the low-energy part of the spectra is shown.}
\end{figure}

In Fig.~\ref{fig:sod1}., the dependence of the resistance
${\cal R}=1/{g}$ on the number
of atoms in the chain is displayed for three different values of $d$.
The result of Lang, who stated that
$R_{N=1}> R_{N=2}$~\cite{LA98}, 
is only found, in our approach, in the case of strong coupling between the chain
and the electrode.
Concerning the coupling strength, there exists a critical value $d_{\rm crit}$
where both
the single atom and the dimer have
approximately the same resistance.
This behaviour can be understood by inspecting the
transmission spectrum, as shown in Fig.~\ref{fig:sod2}.
The value, the linear resistance
of the wire acquires, depends sensitively on the position of the Fermi level
$E_{\mathrm{F}}$ with respect to the modified eigenenergies of the wire.
In order to distinguish between the bare eigenenergies we have displayed
the free wire density of states (DOS) together with the corresponding
$T(E)$ for two different values of the electrode-wire separation.
Intuitively one would expect
that $E_{\mathrm{F}}$ lies in the HOMO-LUMO gap
\index{HOMO-LUMO gap}
for a Na-dimer (the HOMO is twice
occupied) and would almost touch the singly occupied HOMO
in the one atom case. This picture is, however, only exact in the case
of a weak coupling to the electrodes, where the position of
the eigenvalues of the wire remains approximately the same as for an
isolated wire and the broadening induced by the coupling is smaller than
the energy spacing between the eigenvalues.

For $d=6.2 a_{\mathrm{B}}$, however, the eigenstates of Na$_1$ and Na$_2$ are strongly broadened
and shifted by the coupling to the leads. The HOMO and
LUMO (3-fold degenerate) of the single atom
cannot be clearly resolved any more but evolve into a rather broad single
peak. Especially at
$E_{\mathrm{F}}$ the transmission for a single atom becomes smaller than for the
dimer. With increasing distance
the coupling to the electrodes is reduced and thus the renormalization
and broadening of the eigenstates become weaker. At $d=7.0 a_{\mathrm{B}}$, the
HOMO and LUMO of the dimer are already `resolved' and the transmission
$T(E_{\mathrm{F}})$ within the gap is reduced.

In this section, we have introduced as possible bridge molecule a sodium wire. There the simplest assumption has been made for the lead self-energy entering in the calculation, a semi-infinite linear chain with a semi-elliptical spectral density $\Delta$, obtained by Newns~\cite{Newns69}. What are the effects which might arise from using nanoelectrodes such as carbon nanotubes?

\subsection{Focusing on the Leads: Carbon Nanotube Leads}\label{subsect:cnt}
\index{carbon nanotubes}

Carbon nanotube (CNT) conductors have been in the focus of intense experimental and theoretical activity as another promising direction for building blocks of molecular--scale circuits~\cite{KKKCSRO01,RKJTCL00+YWTA01}.
Carbon nanotubes exhibit a wealth of properties depending on their
diameter, on the orientation of graphene roll up, and on their topology, namely
whether they consist of a single cylindrical surface~(single--wall) or many
surfaces~(multi--wall)~\cite{SDD98,McEuen00}.
Carbon nanotubes have been recently used as wiring
elements~\cite{RKJTCL00+YWTA01}, as active
devices~\cite{RKJTCL00+YWTA01,DMAA01+PTYGD01+MSSHA98}, and, attached to
scanning
tunneling microscope (STM) tips, for enhancing their resolution~\cite{WMSS01+WJWCL98}.
With a similar arrangement the fine structure of a twinned DNA molecule has
been
observed~\cite{NKANHYT00}. However, CNT--STM images seem to strongly depend on
the
tip shape and nature of contact with the imaging
substrate~\cite{ANKN01+OBKMRGvdDR00+VdePHMLYMMRF98}.
If carbon nanotubes are attached to other materials to build elements of
molecular circuits, the characterization of
contacts~\cite{TMDSRPNB01+ADX00+dePGWADR99}
becomes again a fundamental issue. This problem arises also when a carbon
nanotube is attached to another
molecular wire, a single molecule or a molecular cluster.

\begin{figure}[t]
\phantom{.}\vspace*{0.1cm}
\centerline{\includegraphics[width=.99\textwidth]{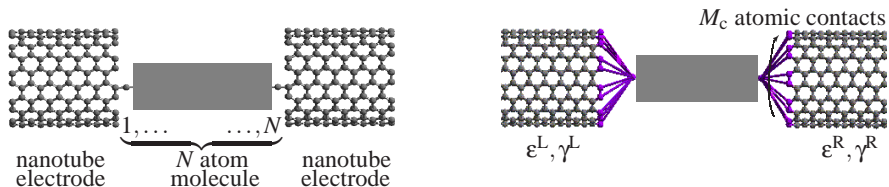}}
\label{page:pippo}
\unitlength1mm
\begin{picture}(0,0)
\put(40,34){\colorbox{white}{\Huge ~~~~~~~}}
\put(40,36){\colorbox{white}{\Huge ~~~~~~~}}
\put(42.4,35){\makebox(0,0)[l]{The role of contacts in molecular electronics~~~~~~~~~\pageref{page:pippo}}}
\end{picture}
\caption{\label{fig:sketch}Schematic representation of the $N$ atom molecule--carbon nanotube hybrid with single (bottom) and multiple (top) contacts.
on--site energies $\varepsilon^{\alpha={\rm L,R}}$ are chosen to be zero.
}
\end{figure}

In this section we present analytic results for the transmission through
a CNT-molecule-CNT system. In particular we analytically derive the spectral density of a single wall armchair carbon nanotubes, needed for calculating the transmission.
A possible configuration is depicted in Fig.~\ref{fig:sketch}.
%%%%%%%%%%%%%%%%%%%%%%%%%%%%%%%%%%%%%%%%%%%%%%%%%%%%%%%%%%%%%%%%%%%%%%%%%%%%%%
%% THE FONTS IN THIS FIGURE HAS BEEN OPTIMIZED FOR DOUBLE COLUMN DISPLAY %%%%%
%%%%%%%%%%%%%%%%%%%%%%%%%%%%%%%%%%%%%%%%%%%%%%%%%%%%%%%%%%%%%%%%%%%%%%%%%%%%%%
\titlerunning{}
%%%%%%%%%%%%%%%%%%%%%%%%%%%%%%%%%%%%%%%%%%%%%%%%%%%%%%%%%%%%%%%%%%%%%%%%%%%%%%
Here a $N$ atom molecule has been adopted as bridge molecule. However, the results we
obtain are valid for any bridge {\it sub conditio} that
the CNTs are contacting the molecular complex only via two {\it
single} atomic contacts (labelled here as $1$ and $N$).
For the system under investigation, where only the two atoms of the
molecule are
coupled to the leads, the formula for the transmission simplifies
to~\cite{CFR01c,YCP02}
\bea
\label{eq:transm-mit-spectral-densities}
T (E) =4 \ \Delta^{\rm L} (E) \Delta^{\rm R} (E) \labs G^{\mathrm{M}}_{1N} \lrb E \rrb \rabs^2 / \mathrm{det} \lrb \vec{Q} \rrb,
\eea
where the spectral densities
$\Delta^{\rm L}$ and $\Delta^{\rm R}$ are the only non-zero elements
$\lrb \vec{\Delta}^{\rm L}\rrb_{11}$ and $\lrb \vec{\Delta}^{\rm R}\rrb_{NN}$,
respectively,
of the matrices $\vec{\Delta}$.
The matrix element $\Delta^{{\rm L (R)}}$ is the spectral density of the
left (right) lead. It is related to the semi-infinite lead Green function
matrix ${\mathcal{G}}^{\rm L(R)}$. It is minus the imaginary part of the
lead self-energies (per spin),
\bea
\label{eq:spectr-density}
\Sigma^{\alpha}
=\Lambda^\alpha - \ii \; \Delta^\alpha
= \sum_{m_{\alpha}, m_{\alpha}^\prime}
\Gamma^{\phantom{*}}_{m^{\phantom{\prime}}_{\alpha}}
\Gamma^*_{m^\prime_{\alpha}}
{\mathcal{G}}^\alpha_{ m_{\alpha}^{\phantom{\prime}} m_{\alpha}^\prime } \, ,
\eea
with $\alpha={\rm L,R}$. Owing to the causality of the self-energy, its real part
$\Lambda$
can be entirely derived from the knowledge of $\Delta$ via a Hilbert
transform. Finally the determinant of $\vec{Q}= \vec{1} -
\vec{\Sigma}\vec{G}^{\mathrm{M}}$ has to be calculated.

The rhs of Eq.~(\ref{eq:transm-mit-spectral-densities}) coincides with
formulas
used to describe electron transfer in molecular systems~\cite{MKR94b,MKR94a}.
The above relationship between the Landauer scattering matrix formalism on the
one side
and transfer Hamiltonian approaches on the other side has been worked out in
the
recent past~\cite{Nitzan01a+HRHS00} showing, {\it de facto}, their equivalence.
This enables us to make use of the formulas from a Bardeen-type picture in
terms of
spectral densities, which is often convenient for an understanding and
analysis of the results obtained.

In calculating the spectral function, we
make use of the assumption of identical left and right leads and drop the
self-energy indices in Eq.~(\ref{eq:spectr-density}).
Since a $\pi$ orbital representation was found to give good agreement with
experiments (even quantitatively)~\cite{SDD98}, the Hamiltonian at hand can be assumed discrete. We can write the lattice
Green function $G = \lrb E +\ii 0^+- \mathbf{H} \rrb^{-1}$
in matrix form by rearranging the two dimensional $n$ lattice coordinate with
honeycomb underlying structure in
the tight-binding Hamiltonian representation. The boundary conditions are
imposed on two cuts parallel to a lattice bond so that the surface of a
semi-infinite CNT contains $2 \ell$ atoms for a so-called $(\ell , \ell)$
armchair CNT.
We assume the $x$ direction to be parallel to the
tubes (and to the transport direction) and $y$ to be the finite transverse
coordinate
(see Fig.~\ref{fig:sketch}.).
The latter is curvilinear with $n_y$ spanning the $2 \ell$ sites with periodic boundary
conditions.

The lattice representation of the lead Green function is needed in the
calculation of
the self-energy contribution.
It can generally be written by projecting the Green operator onto the
localized
state basis,
$\psi_{k_x,k_y} (n_x = {\rm border}, n_y) = \chi_{k_x} \phi_{k_y} (n_y)$,
of the semi-infinite lead:
\bea
\mathcal{G}_{n_y^{\phantom{\prime}} n_y^\prime } \lrb E \rrb &=& \lab
n_y^{\phantom{\prime}} \rabs
\lrb {E + \ii 0^+ -\mathbf{H}} \rrb ^{-1}
\labs n_y^\prime \rab \nonumber \\
\label{eq:lead_gf_general_problem}
&=&\sum_{k_x, k_y} \frac
{ \chi_{k_x} \phi_{k_y} (n_y^{\phantom{\prime}}) \chi^*_{k_x} \phi^*_{k_y}
(n_y^\prime)}
{E + \ii 0^+ - E_{k_x, k_y}} .
\eea
\titlerunning{The role of contacts in molecular electronics}
The eigenvalues of the tight-binding Hamiltonian
\bea
\label{eq:disp_rel_cnt}
E_{\pm} \lrb k_x^j,j\rrb =
\varepsilon
\pm \gamma \sqrt {1 + 4\cos
\lrb \frac{ j
\phantom{k_x^j} \! \! \! \! \!
\pi}{\ell} \rrb \cos\lrb \frac{k_x^j a} 2 \rrb+ 4 \cos^2\lrb \frac{k_x^j a} 2
\rrb} \, ,
\eea
are obtained in a basis set given by symmetric ($+$) and antisymmetric ($-$)
site
configurations of the graphene bipartite lattice, corresponding to
$\pi$ and $\pi^*$ orbitals, respectively~\cite{SFDD92+Wallace47}. The
longitudinal momentum is restricted to the Brillouin zone, $-\pi < k^j_x a <
\pi$, and the transverse wave number $1 \le j \le 2 \ell$ labels $4 \ell$
bands, as many as the number of atoms in the unit cell of a $(\ell , \ell)$
CNT.
The two bands corresponding to $j=\ell$ are singly degenerate. They are
responsible for
the metallic character of armchair carbon nanotubes (these two bands cross at
the
Fermi level $E=\varepsilon$ for $k_x^\ell a = \pm 2 \pi /3$). Also the two
outermost
bands corresponding to $j=2 \ell$ are singly degenerate, while the other
remaining
$(4 \ell -4)$ bands are collected in $(2 \ell -2)$ doubly-degenerate
dispersion curves.

\begin{figure}[t]
\centerline{\includegraphics[width=.85\textwidth]{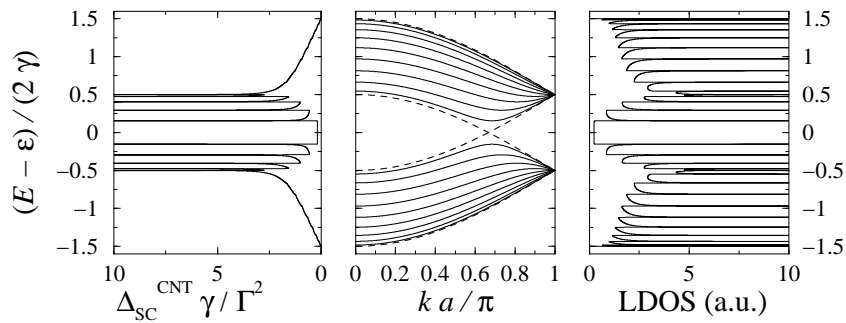}}
\caption{\label{fig:ldos}Left panel: the normalized spectral density for a semi-infinite $(\ell,
\ell)$ CNT lead in
the SC configuration; it corresponds to the LDOS at any atom site at the cut
of the CNT
lead. For comparison the dispersion relation and the
LDOS of an infinite $(\ell, \ell)$ CNT are shown in the middle and right
panel respectively.
Solid lines in the dispersion relation panel indicate doubly degenerate bands,
dashed lines singly degenerate bands.
Here $\ell=10$, and on-site energies and hopping terms refer to $\alpha=\rm
L, R$-leads.}
\end{figure}

The single-particle Green function in a lattice representation
for two sites belonging to the same sub-lattice
can be written as
\bea
\label{eq:general-inf-green-funct-cnt}
\mathcal{G}_{n_y^{\phantom{\prime}} n_y^\prime} \lrb E\rrb
&=&
\frac {a}{2 \pi \ell } \sum_{j, \beta}
\int_{-\pi / a}^{\pi / a} \textrm{d}k^j_x \; \frac {\sin^2 \lrb k^j_x a \rrb
\;
\;
\varphi_j^{\phantom{*}} \lrb n_y^{\phantom{\prime}} \rrb
\varphi_j^* \lrb n_y^\prime \rrb ,
}{E+\ii 0^+ -
E_{\beta} \lrb k_x^j,j\rrb}
\nonumber
\\
&=& \frac{1}{2 \ell} \sum_{j=1}^{2\ell}
\varphi_j^{\phantom{*}} \lrb n^{\phantom{\prime}}_y \rrb
\tilde{G}^j \lrb
E \rrb
\varphi_j^* \lrb n_y^\prime \rrb ,
\eea
where $\varphi_j (n_y) = %
\exp (\ii k_y^j n^{\phantom{j}}_y a)$, with $k_y^j a= \pi j / \ell$, and $1
\le j \le 2 \ell$.
Note that in
Eq.~(\ref{eq:general-inf-green-funct-cnt}), $n_y^{\phantom{\prime}}$ {\it and}
$n_y^\prime$ should be either even or odd (that is they should belong to the
same sublattice).
The semi-infinite longitudinal Green function is given by
\beann
\label{eq:lead_cnt_gf}
\tilde{G}^j \lrb E \rrb =
\frac{a}{8 \pi}
\sum_{\beta=\pm}
\int_{-\pi/a}^{\pi/a}
\textrm{d} k_x^j \;
\frac
{\sin^2 \lrb k_x^j a / 2 \rrb}
{E + \ii 0^+ - E_\beta \lrb k_x^j,j\rrb}
.
\eeann
The integral can be worked out analytically by extending $k_x^j$ to the
complex plane
and adding cross-cancelling paths (parallel to the imaginary axis) along the
semi-infinite
rectangle in the half plane $\Im k_x^j > 0$ and based on the interval
between
$-\pi/a$ and $\pi/a$.
The closing path parallel to the real axis gives a real contribution linear in
energy.
This generalizes the approach by Ferreira \etal~\cite{FDML01},
recently adopted for obtaining an analytical expression for the diagonal Green
function
of infinite achiral tubes, to the case of semi-infinite CNTs.
The determination of the poles inside the integration contour, given by
\beann
- 2 \cos \lrb \frac{q_\beta^j a}{2} \rrb =
\cos \lrb \frac {j \pi}{\ell} \rrb
+ \beta \sqrt{\lrb \frac{E-\varepsilon}{2 \gamma}\rrb^2 - \sin^2
\lrb \frac{j \pi}{\ell} \rrb} ,
\eeann
allows for the calculation of the residues and thus of the surface Green
function.
One finds
\bea
\label{eq:surf_gf_cnt}
\tilde{G}^j \lrb E \rrb =
\frac {1}{2 \gamma} \frac {E -\varepsilon} {2\gamma}
\lrb
1  + \ii
\frac{\ds{
\sin \lrb \frac{q_{\beta_*}^j a}{2} \rrb}}{\ds{
\sqrt{\lrb \frac{E-\varepsilon}{2 \gamma}\rrb^2 - \sin^2
\lrb \frac{j \pi}{\ell} \rrb}
}}
\rrb
\; ,
\eea
where the choice of the contributing pole through the branch parameter
$\beta_* = {\rm sign} \, ( E -\varepsilon )$ has to be
taken into account.
The LDOS, obtained from the imaginary part of the surface Green
function after Eq.~(\ref{eq:surf_gf_cnt}) is plugged into
Eq.~(\ref{eq:general-inf-green-funct-cnt}), is shown in Fig.~\ref{fig:ldos}.
%%%%%%%%%%%%%%%%%%%%%%%%%%%%%%%%%%%%%%%%%%%%%%%%%%%%%%%%%%%%%%%%%%%%%%%%%%%%%%
%% THE FONTS IN THIS FIGURE HAS BEEN OPTIMIZED FOR DOUBLE COLUMN DISPLAY %%%%%
%%%%%%%%%%%%%%%%%%%%%%%%%%%%%%%%%%%%%%%%%%%%%%%%%%%%%%%%%%%%%%%%%%%%%%%%%%%%%%
%%%%%%%%%%%%%%%%%%%%%%%%%%%%%%%%%%%%%%%%%%%%%%%%%%%%%%%%%%%%%%%%%%%%%%%%%%%%%%
It clearly differs from the LDOS of an infinite CNT as depicted
for comparison in the right panel. As for the case of the SLT the pinning of
the
longitudinal wave function at the surface of the semi-infinite systems
{\it cancels} all
border zone anomalies when $q_\pm^j a$ matches multiples of $2 \pi$.
In infinite SLTs these states are the {\it only} resonant states (van Hove
singularities) so
\index{van Hove singularities}
that the surface LDOS of a semi-infinite SLT never diverges. On the contrary, in
CNTs
there are states with zero group velocity outside the border zone which
are responsible for the singularities of the spectral density of
semi-infinite CNTs (left panel of Fig.~\ref{fig:ldos}.).
The self-energy for a CNT lead is more complicate than the one for a SLT
owing to the missing equivalence of the sites belonging to the two different
sub-lattices.
However, since the longitudinal part of the Green function,
Eq.~(\ref{eq:surf_gf_cnt}), is the same for all diagonal and off-diagonal
terms
of the surface Green function,
the self-energy can still be cast into the form
\beann
\Sigma = \frac{1}{2 \ell} \sum_{j=1}^{2 \ell}
\tilde{G}^j ( E )
\eta_{{j} / {\ell}} \lsb \Gamma \rsb .
\eeann
However, for the calculation of
\bea
\label{eq:eta_cnt}
\eta_{{j} / {\ell}} \lsb \Gamma \rsb = \labs
\sum_{m=1}^{2 \ell} \Gamma_m
\varphi_j (m)
\rabs^2 ,
\eea
one has to specify the sub-lattice components of the transverse wave function
and whether they belong to a bonding or anti-bonding molecular state.
Again the distribution of the $\Gamma_m$ contacts is needed in oder to
calculate the
weight $\eta$ and thus the self-energy.
Eq.~(\ref{eq:eta_cnt}) simplifies considerably in the SC case:
$\eta = \Gamma^2$. Since $\eta$ is uniform in $j$,
the self-energy is simply proportional to the diagonal semi-infinite Green
function and, as a consequence, the spectral density is proportional to the
local density of states~(Fig.~\ref{fig:ldos}.).
The MC case ($\Gamma_m=\Gamma_{\rm eff}/\sqrt{2 \ell}$) is also easily
tractable
leading to a sum rule over the possible conducting channels.
However, a direct proof is provided by the intuitive consideration that only
the $\pi$-bonding state can contribute to the MC spectral density
(all the other states have a non-constant spatial modulation provided, \eg, in
Ref.~\cite{CI99}).
Following our notation, the $\pi$-bonding state corresponds to $j=\ell$.
The two
different lead lattice structures carry the same physical information only in
the MC limit case~\cite{CFR01c}.

\subsection{Focusing on the `Molecule Plus Lead' Complex: a Pure Carbon Device}\label{subsect:purecarbon}
\begin{figure}[t]
\centerline{%
\includegraphics[width=.49\textwidth]{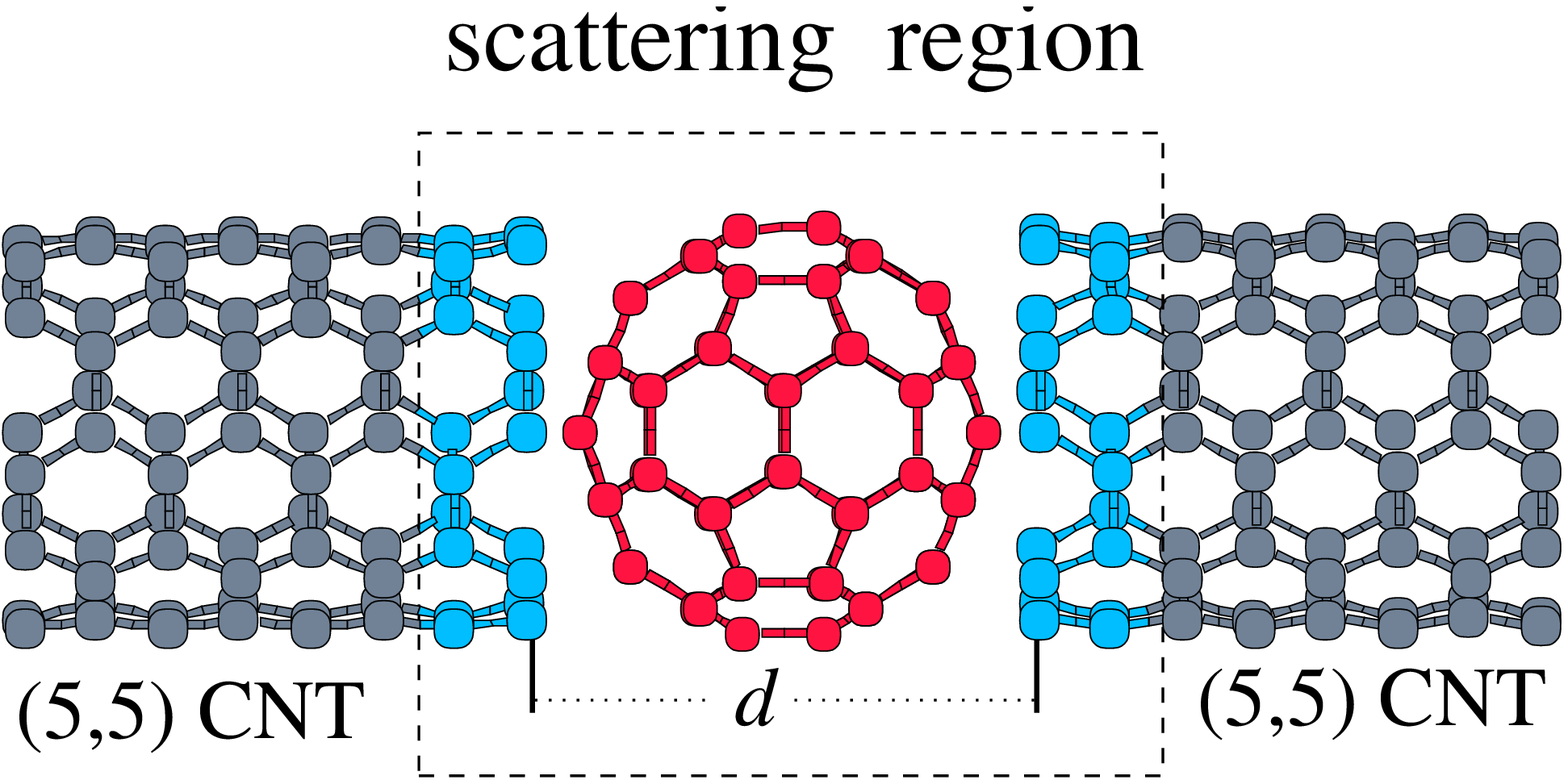}%
\hfill
\includegraphics[width=.48\textwidth]{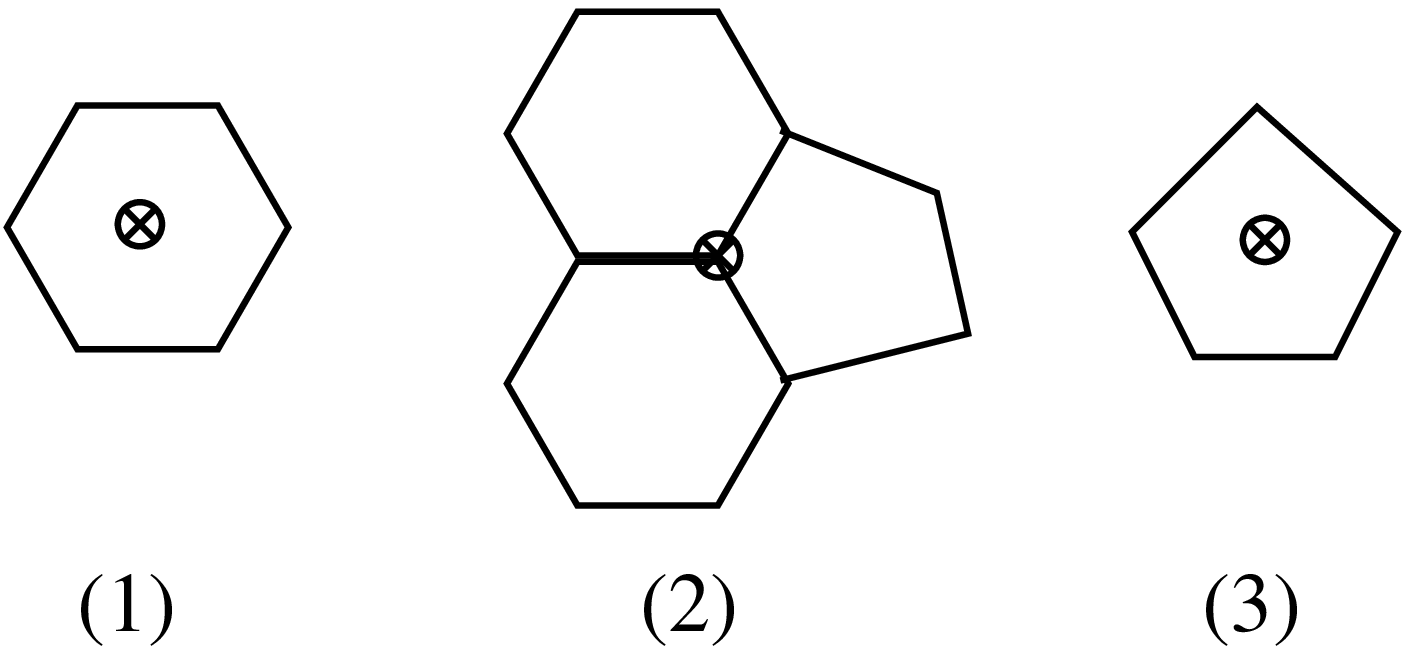}%
}
\caption[]{\label{fig:cntc60}Geometric configuration of the carbon molecular junction.
A C$_{60}$ molecule bridges two (5,5) CNTs.
The right panel represents schematically the different
orientations of C$_{60}$ with respect to the surface cross-sections of the
nanotubes (\eg~the left panel geometry corresponds to
orientation (1)). The nanotube symmetry axis is depicted by a cross inside
a circle.}
\end{figure}

\index{C$_{60}$}
\index{fullerenes}
\index{bucky ball}
In this section we focus on the combination of CNT-leads with a realistic
molecular cluster acting as the central molecule. 
Especially interesting is the case of a monovalent carbon cluster which makes
the system an ``all-carbon'' device.
Therefore, we studied a single C$_{60}$ molecule bridging two single-wall metallic (5,5) carbon nanotubes (CNT). The CNT were
taken symmetric with respect to the plane through the center of
mass of C$_{60}$ and perpendicular to the CNT cylinder axes (see left
panel of Fig.~\ref{fig:cntc60}.).

The central aim of~\cite{GFCGRS02+CGFGRS02} was to exploit the sensitivity of
electron transport to the
topology of the molecule/electrode interface in the
proposed system. In this pure carbon system, charge transfer effects
will be negligible. The Fermi level of the whole system will therefore lie within
\index{HOMO-LUMO gap}
the HOMO-LUMO gap of the isolated C$_{60}$. Therefore, the electronic transport will be mainly
mediated by the overlap of
the tails of the molecular resonances within the HOMO-LUMO gap of C$_{60}$.

\begin{figure}[t]
\centerline{\includegraphics[width=.90\textwidth]{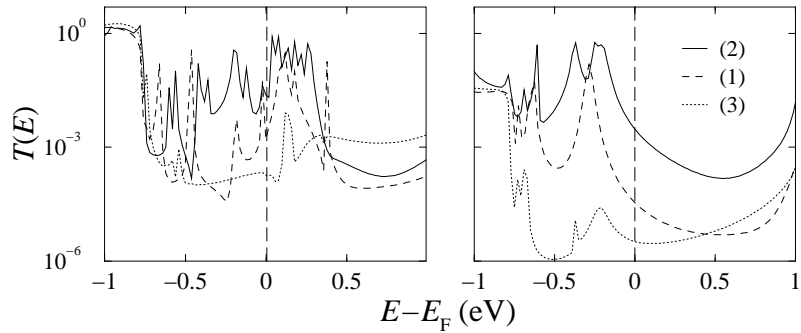}}
\caption[]{\label{fig:relunrel}Transmission results for both unrelaxed (left panel) and relaxed
(right panel) configurations. The tube-tube distance $d$ is fixed at $0.93$ nm. Numbers indicate
different molecular orientations as depicted in the right panel of
Fig.~\ref{fig:cntc60}.}
\end{figure}

The key problem we addressed was
how severely orientational effects do influence the
electronic transport.
To this end several possible orientations of the C$_{60}$ (depicted by the
polygon(s) facing the tube symmetry axis in the right panel of Fig.~\ref{fig:cntc60}.) have been considered for a fixed distance between the molecule and the tubes.
For the sake of comparison, structurally unrelaxed and relaxed molecular junctions were considered.
The basic results are
displayed in Fig.~\ref{fig:relunrel}., for both relaxed and unrelaxed structures.
Surprisingly, at fixed distance, just an atomic scale rotation of
the highly symmetric C$_{60}$ molecule induces a large variation of the
transmission at the Fermi energy by several orders of magnitude.
This can be seen in Fig.~\ref{fig:relunrel}.(right panel) for three different
orientations with maximum, minimum and one intermediate value
of $T(E_{\rm F})$.
As can be seen in Fig.~\ref{fig:relunrel}.(left panel), neglecting relaxation
decisively influences the transmission
properties of the molecular junction. This shows up as a 
different and less smooth behaviour of the transmission. The qualitative difference is related
to the presence of dangling bond states on the CNT surfaces. Such states usually lie
within a gap (a similar situation as that found, \eg, in semiconductor surfaces), in this case
the HOMO-LUMO gap of the isolated molecule. They lead to
the oscillatory behaviour in the transmission for unrelaxed junctions.
Upon relaxation
these states are partly saturated or they rehybridize, moving away from the
middle of the gap. However,
some of them may still lie just above the HOMO or just below the LUMO of C$_{60}$, giving some
contribution to the transmission within the gap.

The results for the relaxed structures reveal that,
at the Fermi energy, the pentagon configuration (3) has a transmission
lower by about \textit{three orders of magnitude} than configuration (2). This fact
could possibly be exploited in an electronic switching device on the
nanoscale, as manipulation of fullerenes by using STM or atomic force
microscope tips is becoming a standard technique in the
field~\cite{ZWWH00}.

\section{Discussion and Conclusions}\label{sect:disc_concl}

\newcommand{\ritem}[1]{\item \dadefinire{#1}}

Summing up our results, we can conclude that modelling transport
at the molecular scale cannot avoid to go through an accurate structure calculation of the
smallest element in the device, namely, the molecular bridge and part of the
attached leads. \textit{Ab initio} methods, although approximated on the
DFT-LDA
level description, provide thus a
fundamental
input to be integrated in standard quantum transport techniques.
The hybrid quantum-transport--\textit{ab-initio} method, reviewed
in Sect.~\ref{sect:method}, provides the necessary playground for a precise
description of linear transport in a two terminal device. The extention to
transistor like configurations goes straightforwardly. On the other hand, the
treatment of the non-equilibrium physics deserves special care, \eg~by using non-equilibrium
Green functions and self-consistency arguments~\cite{XDR01}, which goes far
beyond the limits of this short review.

Two main approximations have been tacitly assumed throughout this work: we
have effectively employed (1) a single electron picture (2) in the limit of
coherent transport.
In doing so, we have been motivated by the fact that small molecules are
typically well adsorbed to the leads providing a \textit{strong
coupling}. Here lies also the main difference with the `artificial
molecules' (obtained by confining the twodimensional electron gas at the
interface of semiconducting heterostructures). In such mesoscopic devices, the electron
states are strongly localized, thus the addition of
any individual electron to the conducting island gives rise to the typical Coulomb
staircase in the device $I$-$V$ characteristics~\cite{GD92-asi}.
Finally, the C$_{60}$ calculations (Par.~\ref{subsect:purecarbon}) have shown that
the molecule clamped between the two mesoscopic leads undergoes to strong
structural modification. Neglecting this effect, \ie, without properly relaxing the
structure, leads to a substantial misestimation
of the linear conductance. The modification of the flexible molecular structure under
the effect of large bias voltages, and a direct computation of charge transfer problems are other important issues which were
left out from the present work (for a recent account in this
direction see~\cite{DVPL02}).

We can then conclude that
different levels of investigation require different theoretical
sophistication. A (semi)empirical description can be extremely useful in
getting the flavor of the most qualitative effects, but can also be
misleading. DFT codes may help in device parameter free calculations and have closer relevance to experiments.
Besides that, it is evident that contacts can change dramatically the conductance profiles, and  further progress in modelling is needed. It might indeed help to separate contact effects from `molecular' effects.
As far as leads are concerned, we have shown that low dimensional leads such as CNT do probe the conductance. Having a reliable model
for the lead self-energy might be the key for cleaning (de-convoluting) spurious
measurements.

Finally, we think that the addition of the richer physical environment that
large molecules do experience
might be easily
scalable on top of the presented transport calculation scheme.
This might be the case for molecular vibrations, and time dependent effects~\cite{CFSK99-epl}.
More difficult would be the extension to comprise electron-electron
interactions~\cite{SCP99,CSK97-epl} and non-equilibrium relaxation.
Attempts to cope with this latter challenge have to include a self-consistent or combined treatment of electronic transport and structural
optimization~\cite{DVPL02}.

\section{Acknowledgments}

B.~C.~M{\"o}bius is gratefully acknowledged for fruitful discussions.
We are indebted to M. Albrecht for providing us with
Fig.~\ref{fig:sodiumsketch}.
GC research at MPI is sponsored by the Schloe{\ss}mann Foundation.
FG and RG gratefully acknowledge financial support by the DFG under FOR
335.

%INDEX%%%%%%%%%%%%%%%%%%%%%%%%%%%%%%%%%%%%%%%%%%%%%%%%%%%%%%%%%%%%%%%
% Please code your entries to include a "mutual" subject index in the
% standard syntax. For your own purposes you may print your
% "personal" index by using the following commands:
%
%\clearpage
%\addcontentsline{toc}{section}{Index}
%\flushbottom
%\printindex
%%%%%%%%%%%%%%%%%%%%%%%%%%%%%%%%%%%%%%%%%%%%%%%%%%%%%%%%%%%%%%%%%%%%%

\end{document}